%% file: moriond.tex
\documentclass[11pt]{article}
\usepackage{moriond,epsfig}
\usepackage{feynmf} 

\def\ZPC{{\em Z. Phys.} C}

\input{physicsShortcuts}

\begin{document}
\begin{fmffile}{moriondDiags}
\vspace*{4cm}
\title{STUDY OF HIGGS BOSON PRODUCTION AT LHC NEAR THE WW RESONANCE}

\author{ C. DELAERE, \\on behalf of the CMS collaboration }

\address{Institut de physique nucl\'eaire, Universit\'e catholique de Louvain,\\ 2 chemnin du cyclotron, 1348 Louvain-la-Neuve, Belgium}

\maketitle\abstracts{
The study of WW Higgs boson decays is one of the key elements of the 
LHC physics program, as these decays are dominant close to the WW resonance. 
Recent results obtained using the full simulation of the detector are 
presented in the framework of the Standard Model.
Direct production, associated WH production and boson fusion 
processes are considered. 
}

\setlength{\unitlength}{1mm}

\input{content.tex}

\end{fmffile}
\end{document}

%% file: physicsShortcuts.tex
\newcommand{\updates}[1]%
        {\fbox{\parbox{\linewidth}{\textbf{Updates from last year:}\\#1}}}

\def\leqsim{\mathbin{\;\raise1pt\hbox{$<$}\kern-8pt\lower3pt\hbox{$\sim$}\;}}
\def\geqsim{\mathbin{\;\raise1pt\hbox{$>$}\kern-8pt\lower3pt\hbox{$\sim$}\;}}
\def\eqref#1{(\ref{#1})}

\newcommand{\Pt}{\ensuremath{P_{t}}}

\newcommand{\gevct}{\ensuremath{{\rm\;GeV}\!/{\rm c}^2}}

\def\ZPC#1#2#3{{\rm Z.~Phys.} {\bf C#1} (#2) #3}

\def\CPC#1#2#3{{\rm Comput.~Phys.~Commun.} {\bf#1} (#2) #3}

%% file: content.tex
\section{Introduction} 

\begin{figure}[hbp!]
\begin{center}
\includegraphics[width=0.49\textwidth]{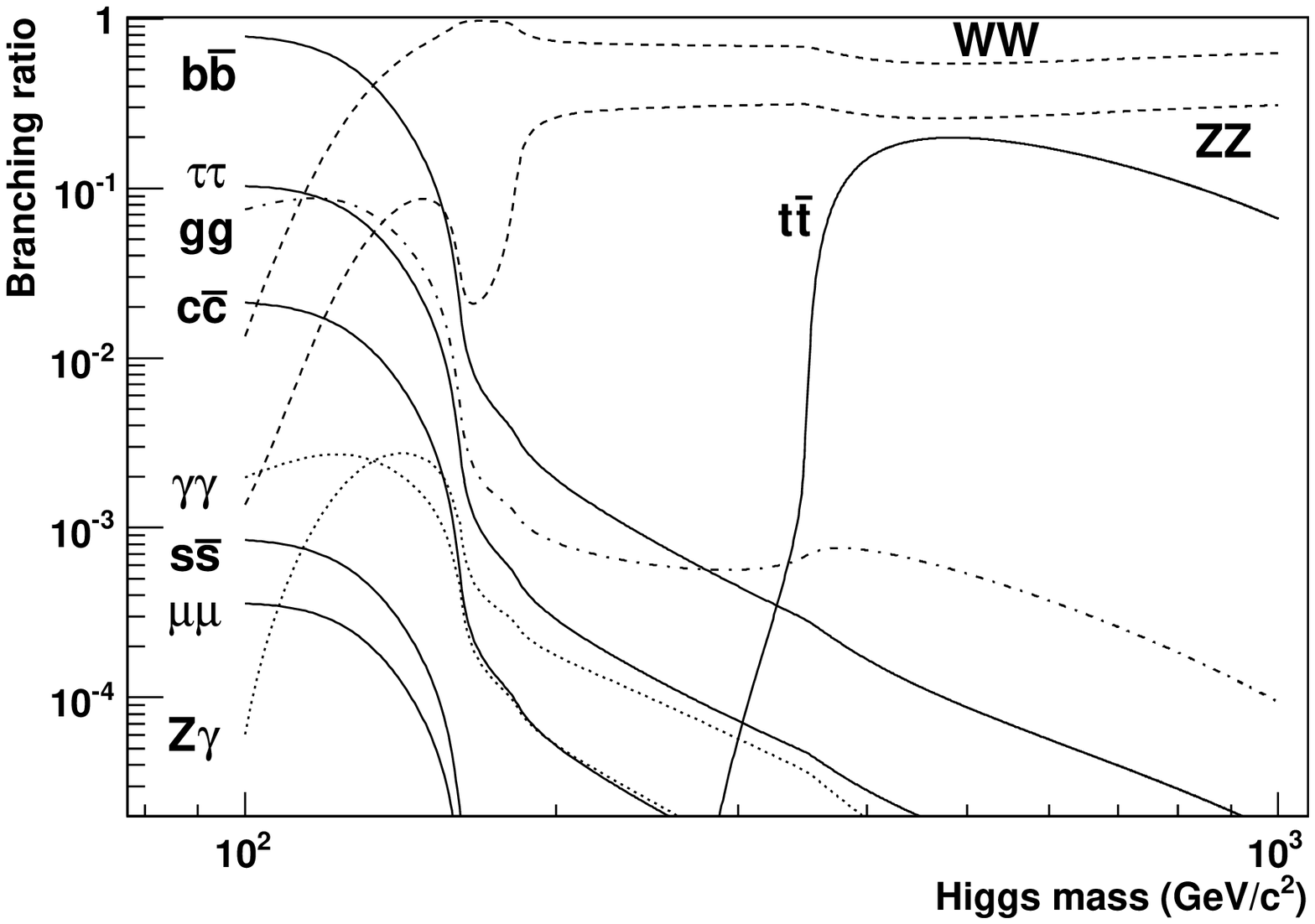}
\hfill
\includegraphics[width=0.49\textwidth]{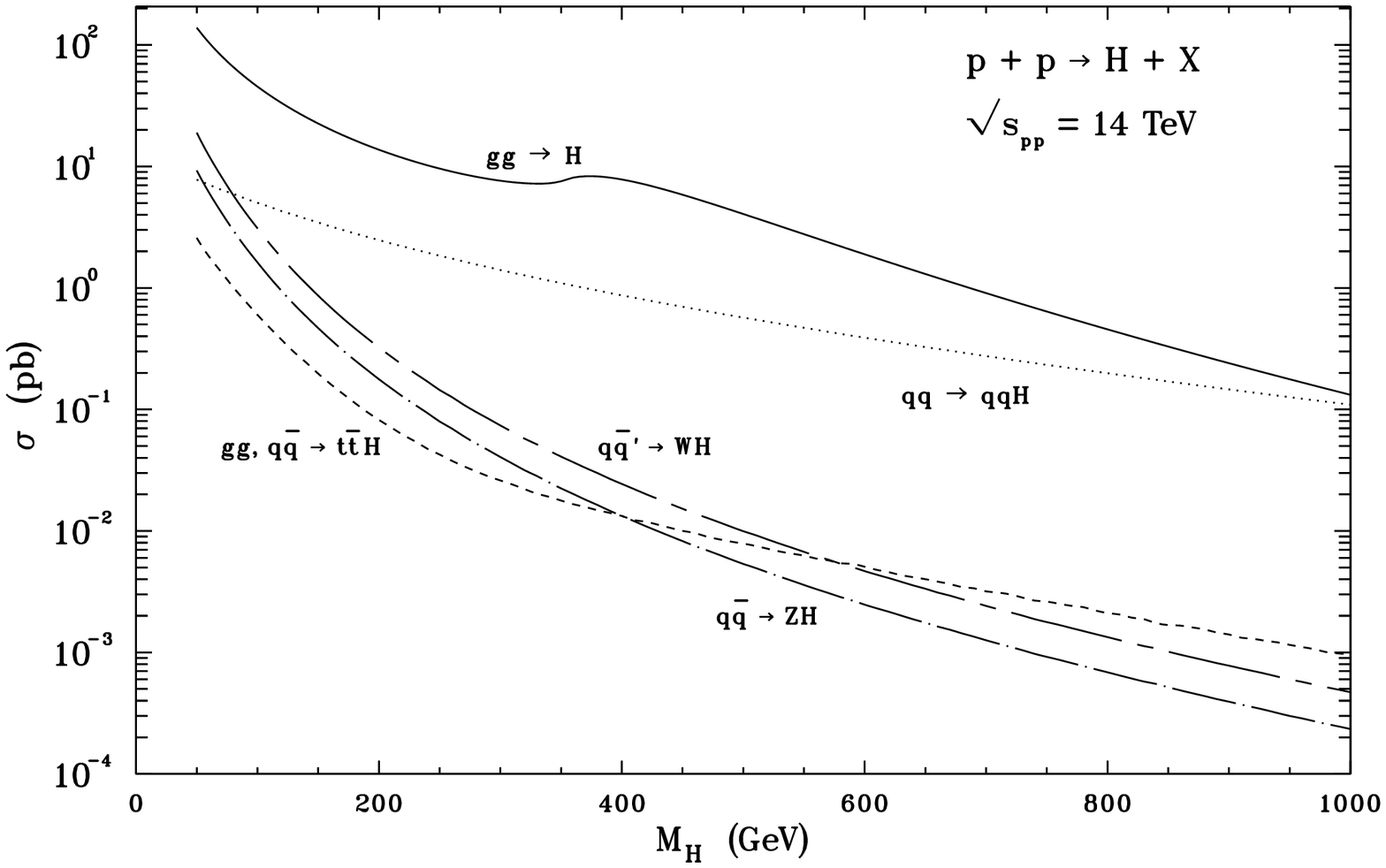}\\
(a) \hskip 0.5\textwidth (b)
\end{center}
\vskip -2ex
\caption{\small 
(a) Next-to-leading order (NLO) calculations for the Standard-Model Higgs boson branching ratios from HDECAY~{\protect\cite{hdecay}};
(b) NLO cross-section calculations for the Standard-Model Higgs boson at LHC~{\protect\cite{kunzt}}.
}
\label{fig:HiggsProdLHCxs}
\label{fig:sxbr}
\end{figure}

Since a few years, various Higgs boson decay channels have been studied in the view of the forthcoming LHC startup. 
The Standard-Model branching ratio of the Higgs boson are presented in Figure~\ref{fig:HiggsProdLHCxs}a.
Close to the WW resonance, all decay modes but the WW decay are suppressed. 
In order to ensure more than a unique observation mode, several production processes must be considered.
The Standard-Model cross-section for each Higgs boson production mechanism is shown in Figure~\ref{fig:HiggsProdLHCxs}b, as a function of the Higgs boson mass.
The dominant contribution to Higgs boson production at LHC comes from gluon-fusion processes, from the weak vector boson fusion mechanism, and from Higgsstrahlung processes, where the Higgs boson is produced in association with a W or Z boson.
All three production mechanisms have been studied by the ATLAS and CMS collaborations. 
In this paper, recent results from the CMS collaboration, obtained using full simulation of the detector, are presented together with results from the ATLAS collaboration, obtained using fast simulation.

\section{WW decays of the Higgs boson}

\subsection{Direct production}

The Higgs boson decay into two W bosons and subsequently into two leptons and two neutrinos ($H \to WW \to \ell\nu\ell\nu$) is expected to be the main discovery channel for the intermediate Higgs-boson mass range, between $\mathrm{2m_W}$ and $\mathrm{2m_Z}$. 
The signature of this decay is characterized by two leptons and high missing energy. 
A study of the discovery potential for this channel, based on a full detector simulation, has been performed by the CMS collaboration~\cite{CMSHWW}. 
In order to get a good NLO estimate for the Higgs-boson production through gluon fusion, the PYTHIA~\cite{pythia} \Pt\ spectrum was reweighted to the MC@NLO~\cite{MCatNLO} prediction, defining \Pt-dependent k-factors. The total cross-section was then scaled to the NLO cross-sections~\cite{NLOrescaling}.
The events generated were passed through a GEANT simulation of CMS. Pile-up corresponding to the LHC low luminosity phase was also generated. The events were then reconstructed using the standard CMS software.
Starting with two oppositely-charged leptons, additional selection cuts were applied in order to reduce the background. A jet veto is applied, and the spin correlation between the two leptons is accounted for by cuts on the angle between the leptons in the transverse plane, and on their invariant mass. There must also be missing transverse momentum.
The cuts were optimized to discover a Higgs boson with a mass between 160 and 170 \gevct.

\begin{figure}
\begin{minipage}[t]{0.48\textwidth}
\begin{center}
\includegraphics[width=\textwidth]{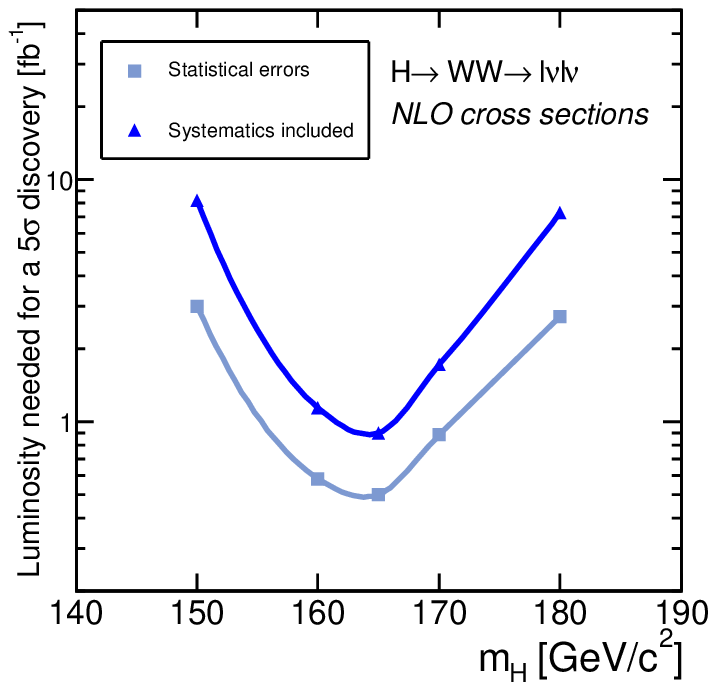}
\put(-35,53){\Large \bf CMS}
\end{center}
\vskip -2ex
\caption{\small 
Luminosity needed for a $5\sigma$ discovery for different Higgs boson masses in the direct production study by CMS.
}
\label{fig:lumi_clb_95_5_direct}
\end{minipage}
\hfill
\begin{minipage}[t]{0.48\textwidth}
\begin{center}
\includegraphics[width=\textwidth]{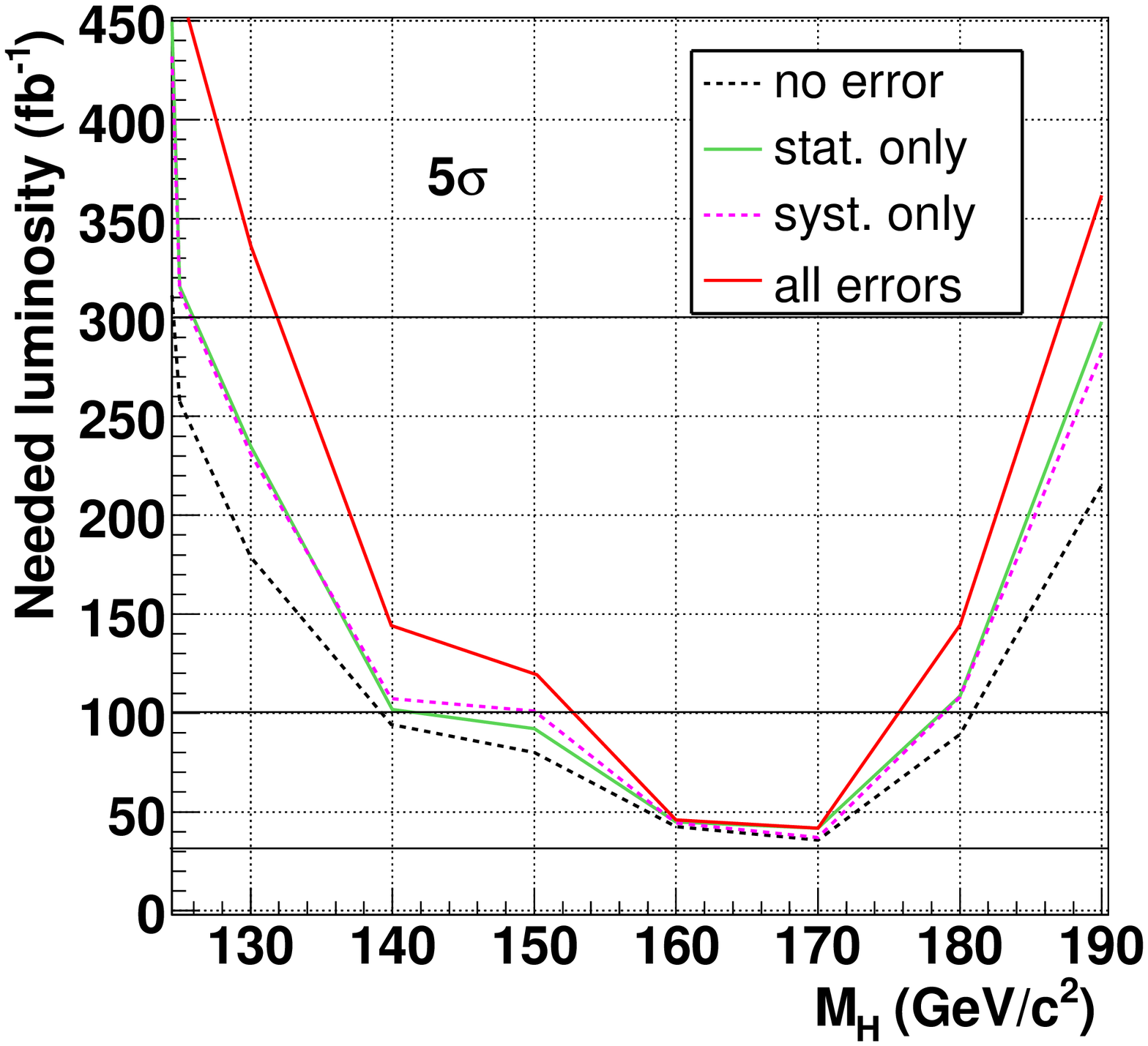}
\put(-50,63){\Large \bf CMS}
\end{center}
\vskip -2ex
\caption{\small Luminosity needed to obtain a $5\sigma$ significance for the associated production channel, with systematics only, with the uncertainty arising from the limited Monte-Carlo statistics only, or with both effects considered.}
\label{fig:lumi_clb_95_5}
\end{minipage}
\end{figure}

Figure~\ref{fig:lumi_clb_95_5_direct} shows the luminosity needed for a $5\sigma$ discovery for different Higgs boson masses.
The signal significance is defined as the probability that the observed background, $\mathrm{N_B}$, fluctuates above the sum of the signal and background, $\mathrm{N_S +N_B}$, following a Poisson distribution with mean $\mu = \mathrm{N_B}$.
A signal with more than $5\sigma$ statistical significance could already be observed with a luminosity of $5\,\mathrm{fb}^{-1}$ for a Higgs-boson mass between 150 and 180 \gevct.

\subsection{Associated production}

Motivations for studying the WH associated production, with a subsequent decay of the Higgs boson into a W pair are twofold.
First, as already mentioned, this channel is one of the few possibilities close to the WW resonance.
Second, the corresponding Feynman diagram contains the $g_{HWW}$ coupling constant twice, that could therefore be precisely measured.
Since three W bosons are produced in this process, the final state is characterized by six fermions in addition to soft remnants from the protons.
The three-lepton channel provides a clean signature and has an interesting signal over background ratio~\cite{CMSWH3l}.

All Standard-Model processes likely to produce three leptons have been considered.
This includes events where three leptons are actually produced but also events with a ``fake'' lepton, a missed lepton, or with the semi-leptonic decay of a B meson.
The production of WWW, WZ, ZZ, $\mathrm{t\bar{t}}$, and Wt have been considered.
Most of the processes used are simulated with PYTHIA, except for WWW, which is generated with CompHep~\cite{comphep}, and Wt generated with TopRex~\cite{toprex}.

In order to select the expected topology, three and only three leptons are required in the event and the total charge of these three leptons is required to be either +1 or -1.
The association of reconstructed leptons either to the Higgs boson decay or to the decay of the W boson that does not come from the Higgs boson is achieved by choosing the two closest opposite-sign leptons.
The third lepton is then supposed to come from the associated W boson.
Most of the background contains a fake lepton or a lepton from the semi-leptonic decay of a B meson, identified in a jet. These are rejected with an isolation criteria. Jet veto and kinematical cuts, like a Z veto, complement the signal selection.

Figure~\ref{fig:lumi_clb_95_5} shows the luminosity needed to obtain a $5\sigma$ significance using this method, with systematics only, with Monte-Carlo statistical uncertainties (for signal and background), or with both effects considered.

\subsection{Boson-fusion processes}

\begin{figure}
\begin{center}
\includegraphics[width=0.48\textwidth]{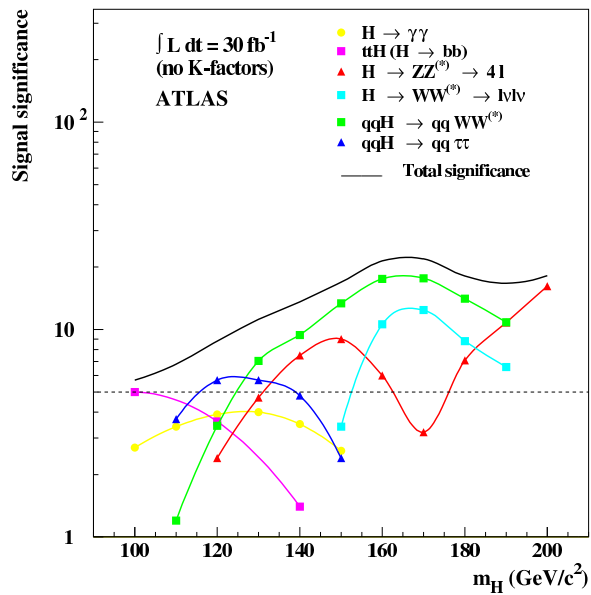}
\end{center}
\vskip -2ex
\caption{\small 
ATLAS sensitivities for the discovery of the Standard Model Higgs boson for an integrated luminosity of $\rm 30\ fb^{-1}$. A systematic uncertainty of 10\% is included for the background in the Weak Boson Fusion channels.
}
\label{fig:ATLASreach}
\end{figure}

When the Higgs boson is produced via boson fusion processes, two jets are expected in the forward and backward regions, with a large rapidity gap with respect to the central event.
By exploiting this feature, an additional background reduction with respect to the direct production can be achieved.
Such studies are being carried on by both collaborations. 
The ATLAS discovery reach in the region of interest presented in Figure~\ref{fig:ATLASreach} shows the appealing potential of such studies.
Using fast simulation of the detector, the work by the ATLAS collaboration~\cite{ATLASWBF} shows that that channel can even be the one with the best significance after $\mathrm{30fb^{-1}}$.
The next steps are to determine the background level from the data, as it is done for other channels, and to estimate the non-trivial systematic uncertainties.

\section{Conclusions}

Close to the WW resonance, all decay modes of the Higgs boson but the WW decay are suppressed. 
In order to ensure more than a unique observation mode, several production processes have been considered by both the CMS and ATLAS collaborations.
This are the gluon fusion, weak vector bosons fusion (VBF) and associated WH production (Higgsstrahlung).
The gluon-fusion mechanism is, with the current state-of-the-art analyses, the main discovery channel. It provides a $5\sigma$ significance for as low as $1\mathrm{fb^{-1}}$ for a 170\gevct\ Higgs boson. 
The Higgsstrahlung mechanism allows to confirm that result, and open one of the only avenues towards the measurement of the coupling of the Higgs boson to W bosons. It is also crucial in ``fermiophobic models''.
Finally, the VBF mechanism shows a very interesting potential, since forward jet tagging and rapidity gap can be used to reduce the background.
It can even be the main channel, if systematics are under control.

\section*{Acknowledgments}
I would like to thank my colleagues from the CMS Higgs group, especially Alexandre Nikitenko for his valuable suggestions, and Anne-Sylvie Giolo-Nicollerat  who provided me with the last results concerning the direct Higgs boson production.Thanks also to Markus Schumacher from the ATLAS collaboration.


%% file: moriond.bbl
\begin{thebibliography}{99}

\bibitem{hdecay}A. Djouadi, J. Kalinowski, M. Spira, ``HDECAY: a Program for Higgs Boson Decays in the Standard Model and its Supersymmetric Extension'', hep-ph/970448 (1997).
\bibitem{kunzt}Z. Kunszt, S. Moretti and W. J. Stirling; \ZPC{74}{479}{1997} and hep-ph/9611397.
\bibitem{CMSHWW}G.Davatz, M.Dittmar,A.-S. Giolo-Nicollerat, Standard Model Higgs Discovery Potential of CMS in H$\to$WW$\to$lnulnu Channel'', CMS NOTE-2006/047.
\bibitem{pythia}T.~Sj\"{o}strand et al., \CPC{135}{2001}{238}.
\bibitem{MCatNLO}S. Frixione and B. R. Webber, ``Matching NLO QCD computations and parton shower simulations'', JHEP0206 (2002) 029 [arXiv:hep-ph/0204244];\\
S. Frixione, P. Nason and B. R. Webber, ``Matching NLO QCD and parton showers in heavy flavour production'', JHEP 0308 (2003) 007 [arXiv:hep-ph/0305252].
\bibitem{NLOrescaling}G. Davatz, G. Dissertori, M. Dittmar, M. Grazzini and F. Pauss, ``Effective K-factors for $gg \to H \to WW \to \ell\nu\ell\nu$  at the LHC'', JHEP 0405 (2004) 009 [arXiv:hep-ph/0402218].
todo: put correct CMS NOTE number
\bibitem{CMSWH3l}C. Delaere, ``Study of associated $\rm WH$ production with
         $\rm H \rightarrow \rm W \rm W^{(\ast)}$
         in the 3 leptons final state.'', CMS NOTE-2006/053.
\bibitem{comphep}E.E. Boos et al., CompHEP: Specialized Package for Automatic Calculations of Elementary Particle Decays and Collisions, SNUTP report 94-116, Seoul, 1994 (hep-ph/9503280).
\bibitem{toprex}S. Slabospitsky and L. Sonnenschein, \CPC{148}{2002}{87}, hep-ph/0201292.
\bibitem{ATLASWBF}S. Asai et al.,``Prospects for the search of a Standard Model Higgs boson in ATLAS using Vector Boson Fusion'', sn-atlas-2003-024.

\end{thebibliography}
